\documentstyle{article}
\input epsf

\newcommand{\beq}{\begin{equation}}
\newcommand{\eeq}{\end{equation}}
\newcommand{\beqa}{\begin{eqnarray}}
\newcommand{\eeqa}{\end{eqnarray}}
\newcommand{\bea}{\begin{eqnarray}}
\newcommand{\eea}{\end{eqnarray}}

\newcommand  {\Mac}      {{\it Macromolecules\ }}

\newcommand  {\COSB}     {{\it Curr.\ Opin.\ Struct.\ Biol.\ }}

\newcommand  {\JCP}      {{\it J.\ Chem.\ Phys.\ }}
\newcommand  {\JMB}      {{\it J.\ Mol.\ Biol.\ }}
\newcommand  {\JP}       {{\it J.\ Phys.\ }}

\newcommand  {\Pro}      {{\it Proteins:\ Struct.\ Funct.\ Genet.\ }}

\newcommand  {\ProSci}   {{\it Protein\ Sci.\ }}

\newcommand  {\PNAS}     {{\it Proc.\ Natl.\ Acad.\ Sci.\ USA\ }}
\newcommand  {\PR}       {{\it Phys.\ Rev.\ }}
\newcommand  {\PRL}      {{\it Phys.\ Rev.\ Lett.\ }}

\newcommand  {\FD}       {{\it Fold.\ Des.\ }}
\input{psfig}
\begin{document}
\begin{titlepage}

\begin{flushright}
LU TP 98-13\\
\today\\
\end{flushright}

\vspace{0.8in}

\LARGE
\begin{center}
{\bf A Minimal Off-Lattice Model\\
for $\alpha$-helical Proteins}\\
\vspace{.3in}
\large
Frank Potthast\footnote{frank@thep.lu.se}\\
\vspace{0.10in}
Complex Systems Group, Department of Theoretical Physics\\ 
University of Lund,  S\"{o}lvegatan 14A,  S-223 62 Lund, Sweden \\
{\tt http://www.thep.lu.se/tf2/complex}\\
\vspace{0.3in}	

Submitted to {\it Journal of Computational Biology}
\end{center}

\normalsize

\vspace{0.4in}
\normalsize
Abstract:\\

A minimal off-lattice model for $\alpha$-helical proteins is presented. 
It is based on hydrophobicity forces and sequence independent local interactions. 
The latter are chosen
so as to favor the formation of $\alpha$-helical structure. They model chirality
and $\alpha$-helical hydrogen bonding. The global structures resulting
from the competition between these forces are studied by means of an efficient
Monte Carlo method. 
The model is tested on two sequences of length $N\!=\!21$ and  $33$ which are intended
to form 2- and 3-helix bundles, respectively. 
%The thermodynamic behavior closely resembles that of the AB model. 
The local structure of 
our model proteins
is compared to that of real $\alpha$-helical proteins, and is found to be very similar.
The two sequences display the desired numbers of helices in the folded phase. Only a few different
relative orientations of the helices are thermodynamically allowed.
Our ability to investigate the thermodynamics relies heavily upon the efficiency of
the used algorithm, simulated tempering; in this Monte Carlo approach, the temperature
becomes a fluctuating variable, enabling the crossing of free-energy barriers.\\


\end{titlepage}
\newpage
\section{Introduction}
Protein folding may be described on all levels of complexity, 
ranging from most simplistic concepts that do not even incorporate the concept of
geometry, to all-atom representations including solvent.
Evidently, different levels of descriptive complexity address
different aspects of the protein folding problem.

Addressing the thermodynamics of a given protein model is generally very difficult due
to the free energy barriers present for compact chains~\cite{Karplus:95}. Therefore, past
thermodynamic studies of proteins have mainly been performed using lattice models.
Lattice models lend themselves to thermodynamic calculations since the conformation space
is discrete, enabling the enumeration of all conformations and therefore the calculation
of exact results for short chains. 
For example, the minimal HP model of Lau and Dill~\cite{Lau:89} has been examined in quite
some detail~\cite{Camacho:93,Dill:95}. 
However, the approximations involved in lattice models are far from well understood~\cite{Erik},
and their geometry is clearly not that of real proteins. Furthermore, the energy barriers between
the states may be poorly represented due to the discreteness of the conformation space.
 
In this paper, we study the thermodynamic behavior of a simple off-lattice model for protein folding.
Studies of similar, minimal, off-lattice  models have been performed before, 
see e.g.~\cite{Iori,Fukugita:93,Veitshans,Stillinger,Irback:95,Irback:97}.
The main difference between our model and those studied previously is that it is deliberately constructed so as to model
the $C_{\alpha}$ backbone geometry of $\alpha$-helical proteins. A somewhat similar approach was taken in~\cite{Rey}. 
However, these authors used a conventional Monte Carlo method, and 
reliable thermodynamic averages for global quantities were not obtained.

Our starting point is the model recently suggested by Irb\"ack et al.~\cite{Irback:97}.
The major change is a modification of the local interactions which we here choose so as to model  
$\alpha$-helical structure. We stick to the original concept of presenting each
amino acid as a single site. Furthermore, only two types of residues are considered, hydrophobic and
hydrophilic. The residues are linked by rigid bonds.
The local interactions in our model are meant to 
model the chirality of the amino acids and the hydrogen bonds found in real $\alpha$-helices.

It has been shown by Kamtekar et al.~\cite{Kamtekar} that the binary sequence pattern 
of hydrophobic and hydrophilic amino acids is of central importance in 
the design of de novo proteins.
%; the relatively simple information
%in the binary code is sufficient for the design of enormous numbers of de novo proteins that
%fold into compact $\alpha$-helical structures. 
The model presented here
has some potential to be of help when pursuing this 
type of protein design. 
In particular, it may give valuable hints on different thermodynamically stable structures of a given
hydrophobic/hydrophilic sequence; such information would certainly be useful in this design
approach.

\section{Methods}
\subsection{The Model}

Our starting point is the model proposed in~\cite{Irback:97}, to be referred to as the
AB model, which is briefly described in the appendix.
In the AB model, there are sequences which 
are thermodynamically stable at kinetically acceptable temperatures~\cite{Irback:97}. Also, there are
strong regularities in the local structure of the chains, qualitatively similar to those 
for real proteins. 
However, at a more quantitative level, the local  
structure observed in~\cite{Irback:97} is very different from that
of real proteins - partly because it is insensitive to space reflections (no chirality).
The modifications of the AB model presented in this paper are meant 
to model $\alpha$-helical local structure; in particular, chirality is introduced.
Globally, the modifications lead to sizes similar to those of real proteins. 

Let us call the right-handed $\alpha$-helical structure found in real proteins
an {\bf ideal $\alpha$-helix};  this helix has 3.6 amino acids
per turn and a translation of 5.4$\mbox{\AA}$ per turn. An ideal $\alpha$-helix is
completely described by the geometry of three successive (virtual)
$C_{\alpha}$-$C_{\alpha}$ bonds; that is by one torsional ($\alpha_i$) 
and two bend angles ($\tau_i,\tau_{i+1}$).
Equivalently, one may describe an ideal $\alpha$-helix by the
mutual distances $r^{\alpha}_{ij}$  of the four participating $C_{\alpha}$ atoms (plus
the correct chirality). 
%Before we start to define the model,
%let us make a technical remark: In this paper, lengths will repetively be given in units of
%$1 \mbox{L}\!\equiv\!3.8 \mbox{{\AA}}$, corresponding to 
%the $C_\alpha_i \!-\!C_\alpha_{i+1}$ distance in the trans-conformation of the polypeptide backbone.\\

In our model, each residue is represented by a single site corresponding to the
$C_{\alpha}$ position in the polypeptide backbone. 
These sites are linked by rigid bonds, $\vec{b}_i$, of length $3.8\mbox{\AA}$. 
The shape and energy of an $N$-mer is specified by  the $N\!-\!1$  bond vectors $\vec{b}_i$. 
$r_{ij}$ denotes the distance between residues $i$
and $j$. $\sigma_1,\ldots,\sigma_N$ is a binary string that 
specifies the primary sequence; we consider hydrophobic ($\sigma\!=\!0$) and 
hydrophilic ($\sigma\!=\!1$) monomers only. The energy function is defined to be
\beq
E(\vec{b}; \sigma) =  
\sum_{i=1}^{N-2}\sum_{j=i+2}^N 
4\epsilon(\sigma_i,\sigma_j)\left(  \left(\frac{{A_{ij}}}{{r_{ij}}}\right)   ^{12}-
\left(   \frac{A_{ij}}{r_{ij}}   \right)^{6}\right)\ + \sum_{i=1}^{N-3}E_l(i) 
\label{energy_definition}
\eeq
The first term in this equation consists of Lennard-Jones (LJ) interactions.
The LJ parameters $A_{ij}$ are chosen as 
$A_{ij}=2^{-1/6}r_{ij}^\alpha$ for $\mid\!j\!-\!i\!\mid \leq 4$, 
where   $r_{ij}^\alpha$ denotes the distance in an ideal $\alpha$-helix.
This choice of $A_{ij}$ gives the corresponding
terms in eq.~\ref{energy_definition} a minimum at $r_{ij}^{\alpha}$.
For all other ($i,j$), we set $A_{ij}\!\approx\!6.01\mbox{{\AA}}$, 
which is calculated from the average volume,
$161 {\mbox{\AA}^3}$, of an amino acid~\cite{Creighton:proteins}. Hence there
are four different $A_{ij}$ values, three of which are determined by the 
geometry of an ideal $\alpha$-helix, the fourth being defined by the average size
of an amino acid. Finally,
the depth of the LJ potential is chosen so as to favor the formation of a hydrophobic core;
$\epsilon(0,0)\!=\!1$ for a hydrophobic-hydrophobic pair,  $\epsilon\!=\!1/2$ for all 
other pairs. 

\begin{figure}
\mbox{
\epsfxsize=4.5cm
\epsfbox{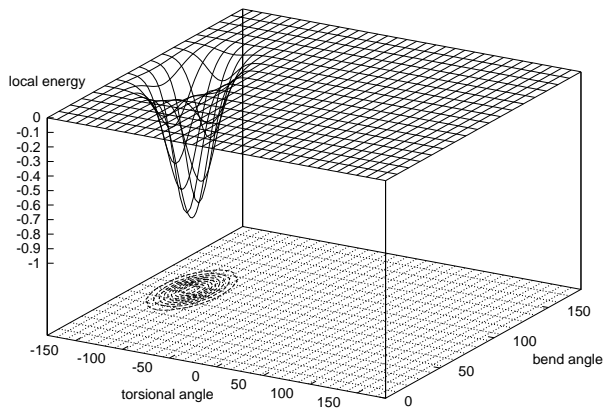}
}
\caption{Local energy $E_l(i)$ (Eq.~\ref{local_equation})
displayed in one bond and one torsional angle. The second
bend angle is set to $\hat{\tau}$, the bend angle of an ideal $\alpha$-helix. $E_l(i)$
has a single minimum of depth one around the $\alpha$-helical conformation and is close to zero elsewhere.
Equipotential lines for $E_l(i)\in\{-0.1,-0.2,...,-0.9\}$ are displayed with circles.
}
\label{local_energy_plot}
\end{figure}

The sequence-independent local interaction 
$E_l(i)$ depends on the  torsional and bend angles 
of three successive bond vectors ($\alpha_i,\tau_i,\tau_{i+1}$), and is 
given by a negative gaussian with unit depth and
width $w$;
\beq
E_l(i)=- \exp\Big( -\! \frac{(\alpha_i\!-\!\hat{\alpha})^2\!+\!(\tau_i\!-\!\hat{\tau})^2\!
+\!(\tau_{i+1}\!-\!\hat{\tau})^2}{2 w^2}\Big)
\;\;\;\;\;\;E_l(i)\in [-1,0]
\label{local_equation}
\eeq
where $\hat{\alpha}$ and $\hat{\tau}$ are the torsional and bend angles in an ideal $\alpha$-helix. 
$E_l(i)$ is close to minus one if and only if the
three participating bond vectors are close to $\alpha$-helical structure, which is illustrated in Fig.~\ref{local_energy_plot}.
The width is taken to be $w\!=\!1/4\;\mbox{rad}\!\approx\!14.32^{\circ}$; this choice is
justified by the resulting angle distributions, which will be presented in 
section 3.1.
The aim of the local interaction is twofold. First, it introduces a 
right-handed chirality into the model that is not
present in the LJ term alone. Second, it is meant to model the hydrogen bonds of 
the $\alpha$-helical conformation. Observe that the depth of the 
local interaction $E_l(i)$ equals the depth of a hydrophobic-hydrophobic contact.
Exact values of the parameters and technicalities, as well as a comparison to the
AB model, can be found in the appendix.

In section 3.1, it will be shown that, not surprisingly, the model indeed gives rise to $\alpha$-helical structure. 
This is done by comparing the
local structure of our model to that of real $\alpha$-helical proteins.
Quantitatively very similar results are found.
Furthermore, we examine 
the overall topology of the simulated sequences.
As will be shown in section 3.2, we find topologies corresponding to 2- and 3-helix bundles.
By looking at the $C_{\alpha}$-$C_{\alpha}$ distance distribution
functions, it can be seen that even the linear dimensions of our model
proteins are similar to those of real ones. 

\subsection{Sequences}
We have studied the behavior of two sequences in this model.
These were deliberately chosen so as to be consistent with $\alpha$-helical
structure. Specifically, we start from a sequence segment of length 14,
which is known to occur in amphiphilic $\alpha$-helices~\cite{West}.
This segment has been used earlier for the design of (real) de novo $\alpha$-helical
proteins~\cite{Kamtekar}.
We take the first 9 positions of this $N\!=\!14$ segment as our basic building block.
Our two sequences are then obtained by taking two (three) copies of this building block,
and connecting them by hydrophilic segments of length three.
These sequences have length $N\!=\!21$ and $33$  respectively, and are 
given in table~\ref{3sequences}, where the spaces are meant to clarify the construction of the sequences.
It should be obvious that these are meant as templates for 
2- and 3-helix bundles, respectively.  
\begin{table}[htp]
\begin{center}
\begin{tabular}{c|c}
Length $N$  & sequence $\sigma_1,...,\sigma_N$ \\ \hline
21          & 101100110 111 101100110                                 \\
33          & 101100110 111 101100110 111 101100110                   \\
\end{tabular}
\end{center}
\caption{The two sequences studied. Basic building blocks of length 9 are connected by
hydrophilic segments of length 3.}
\label{3sequences}
\end{table}

\subsection{Monte Carlo Methods: Simulated Tempering}
We calculate thermodynamic properties by using the method of simulated tempering~\cite{Ryssland,Marinari:92,
Irback:95}. To simulate proteins is notoriously difficult, due to the presence of a
rugged energy landscape. One tries to overcome this problem in simulated tempering by
treating the temperature as a fluctuating variable.
This means that one simulates a joint distribution in conformation and temperature,
which is taken to be
\beq
P(\vec{b},k)\propto\exp(-g_k-E(\vec{b},\sigma)/T_k) \,,
\eeq
where $T_k$, $k=1,\ldots,K$, are the allowed values of the 
temperature. 
The $g_k$'s are free parameters that determine the probabilities $P(T_k)$ of visiting
the different temperatures. Although the method is free from systematic errors for any 
set of $g_k$, it is crucial for the performance of the algorithm to make a careful choice
of these parameters. 
In our simulations,  the parameters $g_k$ were adjusted so as to have a roughly
uniform distribution $P(T_k)$.
This was done by means of trial runs. Details may be found in~\cite{Irback:95}. 

The simulations reported in this paper were carried out using a set of $K\!=\!20$ 
temperatures, ranging from $T_1\!=\!0.15$ to $T_{20}\!=\!1.0$ with $1/T_{i}-1/T_{i+1}$ constant.
On a DEC ALPHA 200 (266 MHz), they took
around 30 hours for $N\!=\!21$ and around 500 hours for $N\!=\!33$.
In both cases, around 20 percent of the total computing time was spent on tuning
of the weights $g_k$.
We also simulated a sequence of length $N\!=\!45$, whose construction is
analogous to that of the other two sequences.
We spent 1000 CPU hours on this sequence,
and believe that we obtained reliable estimates for its local properties. However, the results for global
properties must be interpreted with care for this sequence.
By contrast,
we feel very confident that the simulations for $N\!=\!21$ and  $N\!=\!33$ 
are under control.

\section{Results}
The overall thermodynamic behavior of our model turns out to closely resemble
that of the AB model~\cite{Irback:97}. In particular, the chains compactify gradually with 
decreasing temperature, and the folding transition takes place in the compact phase. 
Furthermore, we do not observe bimodal distributions in energy, compactness or the 
local interactions $E_l(i)$. All these observations are in agreement with earlier results
for the AB model. By contrast, our model differs significantly from the AB model when it
comes to structural properties. We shall therefore focus on these. Local structure as well as
global topology and stability will be discussed in some detail. 

\subsection{Local Structure: Comparison with Real Proteins}
%Local structures of real proteins are well-documented, they are included here for illustrational
%purposes and for comparison with our results.
In this section we examine the local structure of the chains in our model, and compare it to that
of real $\alpha$-helical proteins. The protein data were extracted using the $C_{\alpha}$
coordinates for 27 proteins
containing mainly $\alpha$-helices and almost no
$\beta$-sheets. All these 27 proteins are listed in the all-$\alpha$-helical class of the SCOP 
database~\cite{scop}.
The corresponding structures were taken from the Brookhaven Protein
Data Bank (PDB)~\cite{lo_Bernstein:77} \footnote{The PDB accession codes are:
3sdh, 1ctj, 1enh, 2erl, 2end, 1lis, 1hme, 1nfn, 1bcf, 1rhg, 1acp, 1rop, 1coo, 4icb,
1utg, 1fia, 2wrp, 2tct, 1fps, 1ecm, 1aep, 1axn, 1hiw, 2abk, 1eci, 2abd, 1c5a.}.
One of these proteins, the $N\!=\!54$ DNA-binding protein 1enh, 
which  contains three $\alpha$-helices~\cite{KISSINGER}, will be considered in more detail.

The local structure will be probed in two ways,
by bend and torsional angle distributions and by bond-bond correlations. 

\begin{figure}
\mbox{
\epsfxsize=8cm
\epsfbox{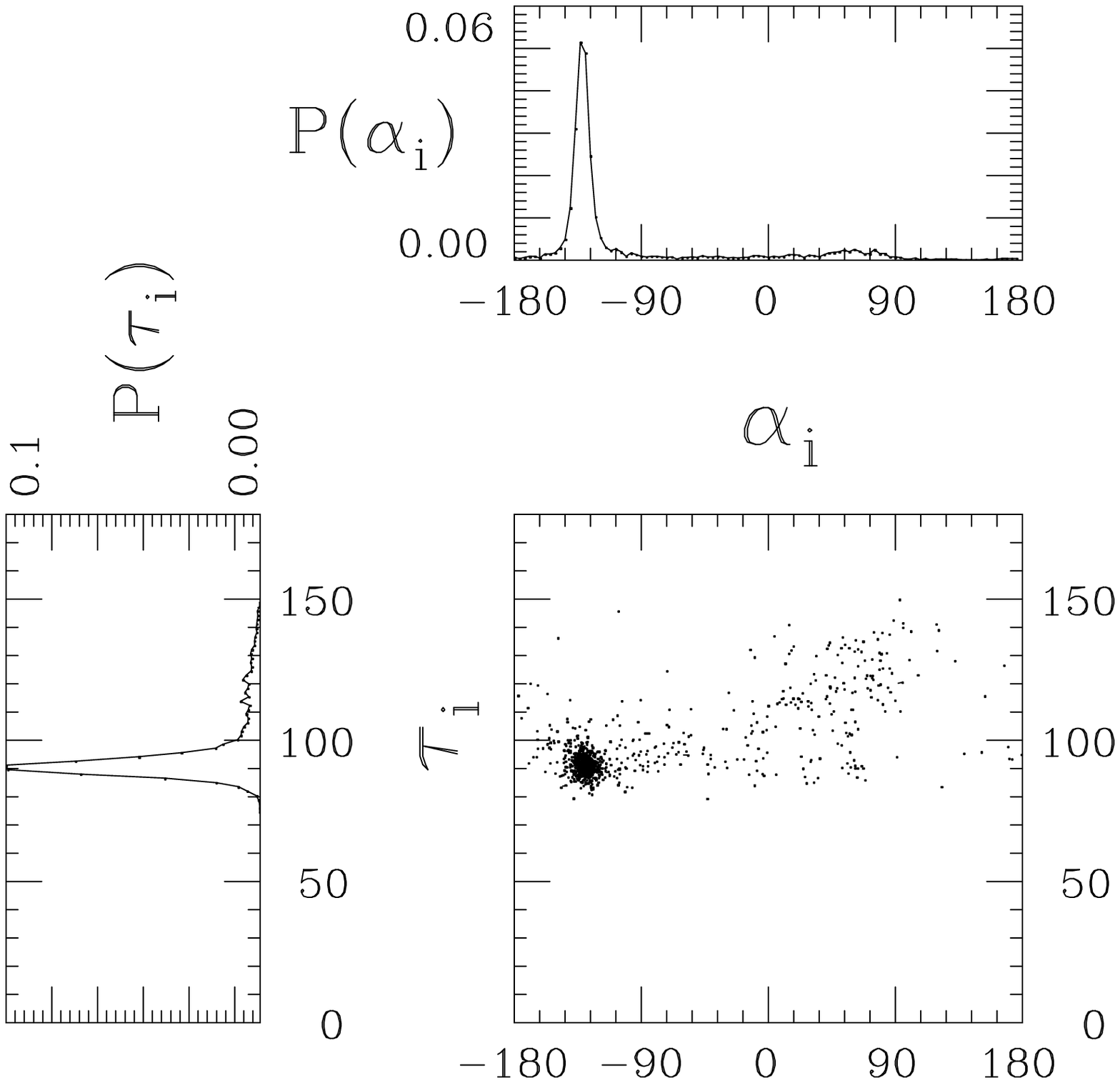}
\epsfxsize=8cm
\epsfbox{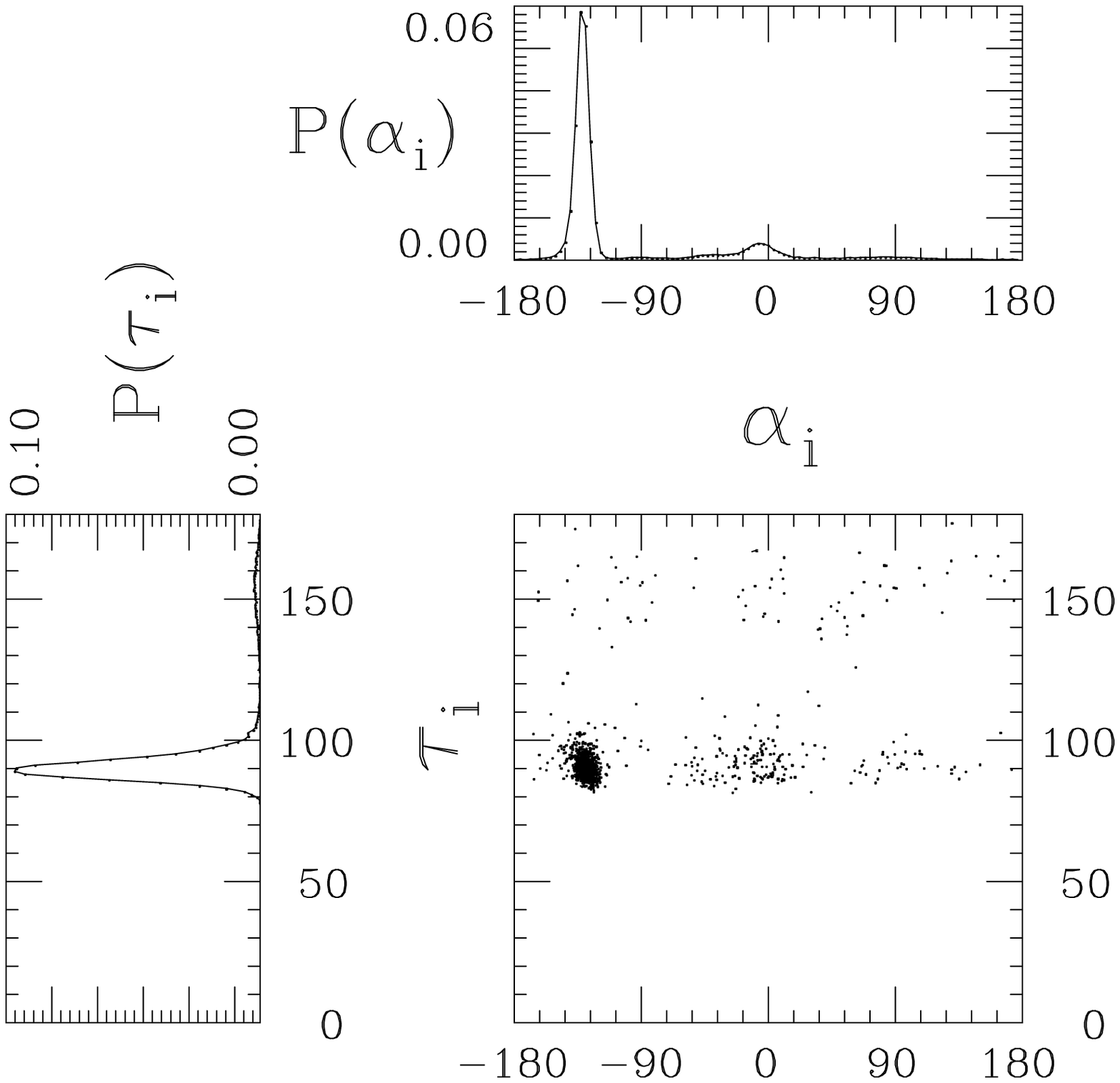}
}
\vspace{-1.5cm}
\caption{
Bond ($\tau_i$) and torsional ($\alpha_i$) angle distributions.
Left: For the 27 $\alpha$-helical proteins taken from the PDB database. 
Right: Model results for the $N\!=\!21$ sequence at $T\!=\!0.15$.
The other two sequences simulated, with $N\!=33\!$ and $N\!=\!45$, give very similar distributions. 
Both scatter plots contain 1000 data points.
}
\label{angle_results}
\end{figure}

In Fig.~\ref{angle_results}, left plot, bend and torsional angle distributions are displayed for the
set of 27 $\alpha$-helical proteins. The dominance of $\alpha$-helical structure
manifests itself in the pronounced peak centered at ${\alpha}\approx-129^\circ$ and $\tau\approx 90^\circ$.
The same distributions, for our $N\!=\!21$ sequence, are plotted on the right side of Fig.~\ref{angle_results}.
The results for the other two simulated sequences  are very similar.
This is in line with the behavior of the AB model,  where the angle distributions
are fairly sequence-independent as well~\cite{Irback:97}. Not surprisingly, we find 
that there is indeed a peak in the distribution
centered at the position of the ideal $\alpha$-helix. 
This peak contains roughly the same amount of probability as for real
$\alpha$-helical proteins.
Moreover, the width of this peak is very similar
to that of the experimental structure, which justifies our ad hoc value for the
width $w$ of the exponential in Eq.~\ref{local_equation}.
Another and much weaker peak in the angle distribution for our
model is found at $(\alpha,\tau)$=$(0^\circ,90^\circ)$, 
which is an artifact of our model not present in real proteins.

\begin{figure}[tbp]
\begin{center}
\vspace{-42mm}
\mbox{\hspace{-31mm}\psfig{figure=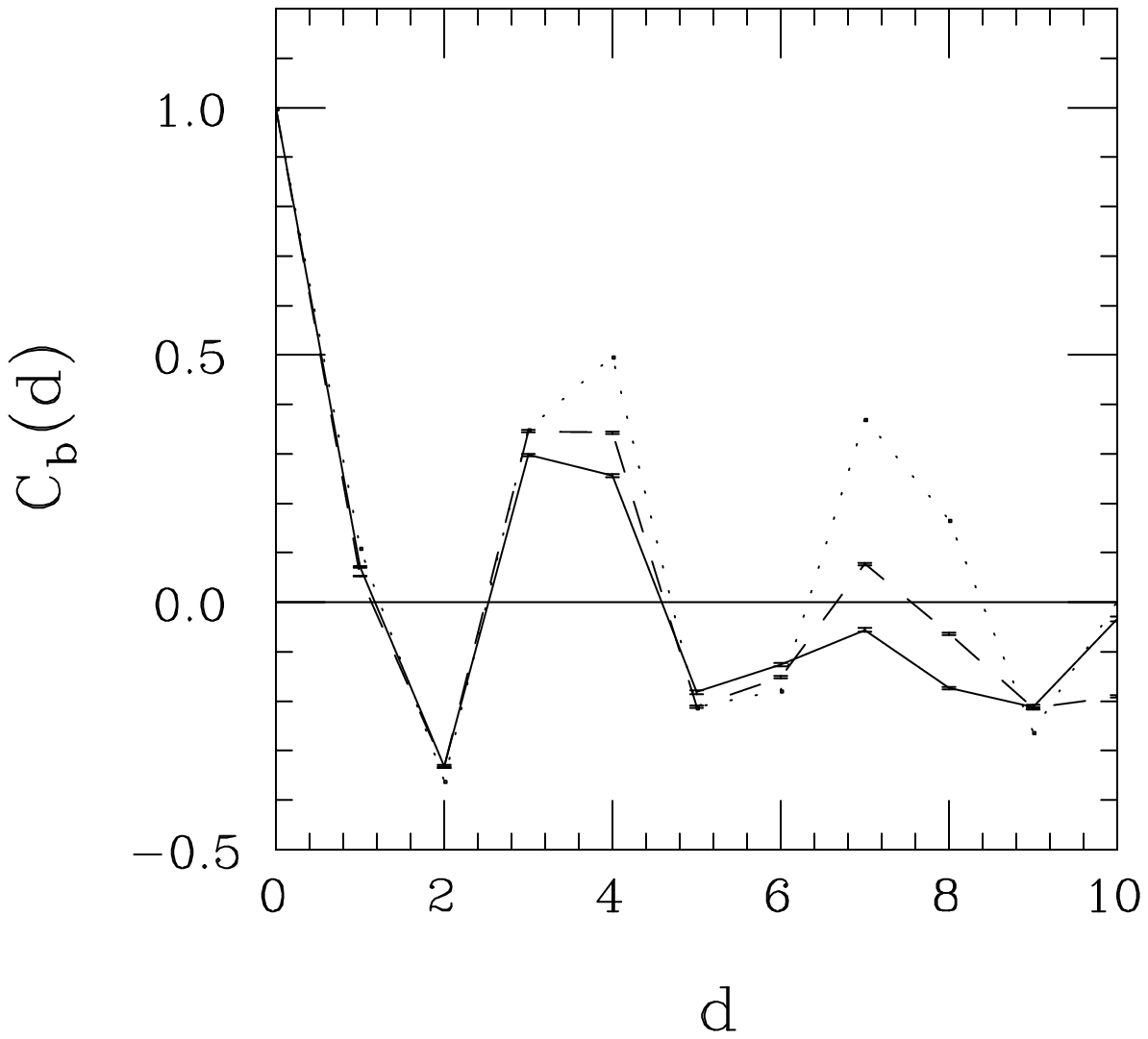,width=10.5cm,height=14cm}
\hspace{-30mm}\psfig{figure=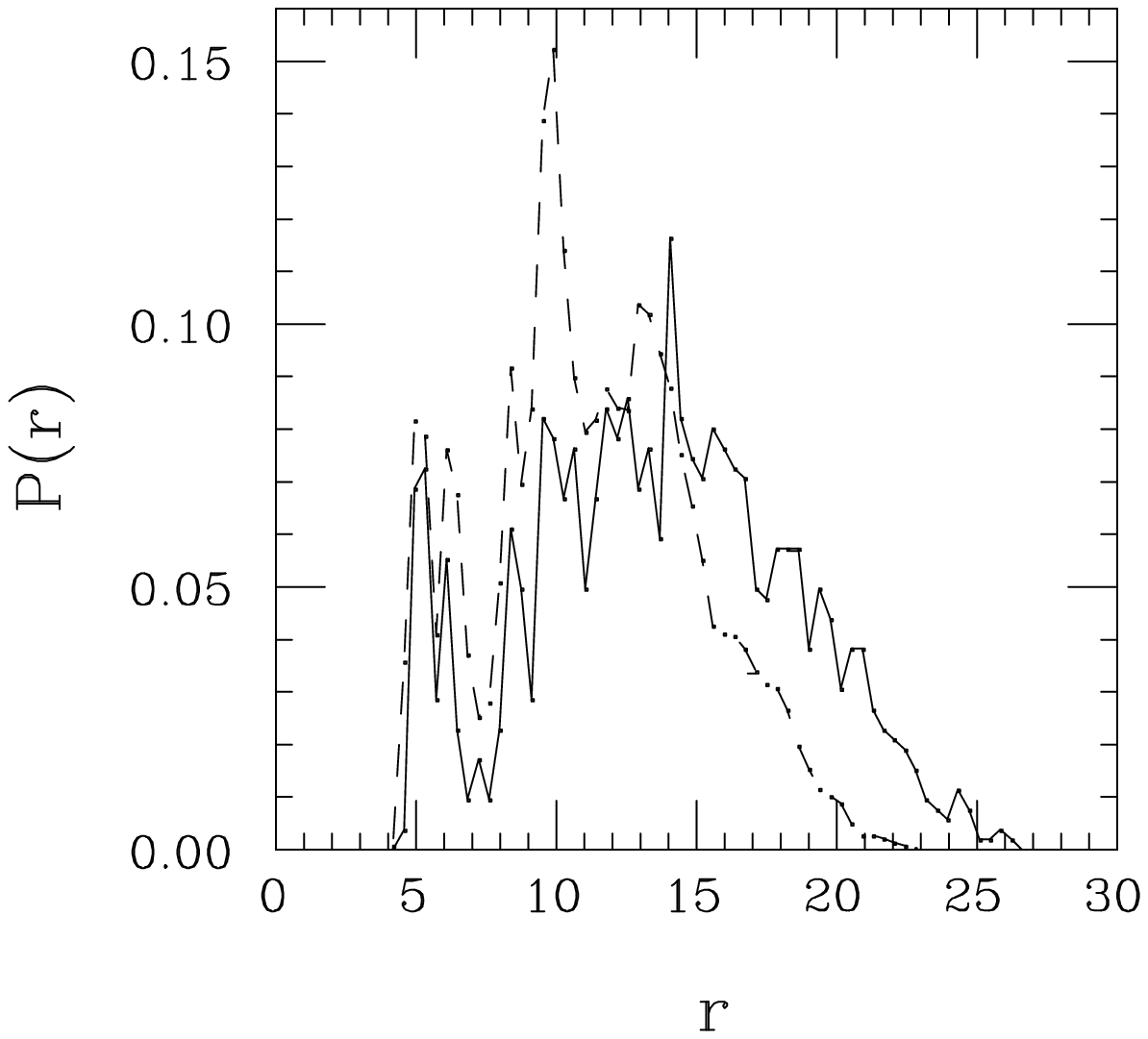,width=10.5cm,height=14cm}}
\vspace{-42mm}
\end{center}
\caption{
Left: Correlation functions $C_b(d)$ for our
model at $T=0.15$ for $N\!=\!21$ (solid line) and $N\!=\!33$ (dashed line).
$C_b(d)$ for the 27 $\alpha$-helical proteins listed in the text
is given by the dotted line.
Right: $C_{\alpha}$-$C_{\alpha}$ distance distribution $P(r)$ 
for the (molecular dynamics refined) experimental structure of the 
DNA binding protein 1enh ($N\!=\!54$, solid line),
and  our model ($N\!=\!45$, dashed line) at $T\!=\!0.15$. Nearest-neighbor
pairs are  not included. 
}
\label{lo_fprot_corr}
\end{figure}

Our second way to monitor local structure are the bond-bond correlations, which  are studied using the function 
\beq
%C_b(d)={1\over N-d-1}\sum_{i=1}^{N-d-1}\frac{ \vec{b}_i\cdot\vec{b}_{i+d}}{\mid\!\vec{b}_i\!\mid \mid\!\vec{b}_{i+d}\!\mid}
C_b(d)=   \left\langle \frac{ \vec{b}_i\cdot\vec{b}_{i+d}}{\mid\!\vec{b}_i\!\mid \mid\!\vec{b}_{i+d}\!\mid}\right\rangle
\;\;\;\;\;\;\;
C_b(d)\in[-1,1],
\label{lo_corr}
\eeq
where $\langle\cdot\rangle$ denotes the average over all positions $i$ and observed structures (with $d$ fixed).
$C_b(d)$ is a measure of the average alignment of bond vectors at a given topological distance $d$
along the chain.
It is normalized so that $C_b(0)=1$.
In Fig.~\ref{lo_fprot_corr} (dotted line, left plot),
we  show the correlation function $C_b(d)$ for the 27 $\alpha$-helical proteins. As can be seen
from this figure, there are significant correlations at least out to separations of  
about ten residues. The oscillations can be related to the presence of 
$\alpha$-helical structure, which has a period of 3.6.      
In \cite{Irback:97} it was shown that the AB model gives roughly the right 
correlation length, but fails to
reproduce the periodicity of 3.6.
$C_b(d)$ for our model is shown in Fig.~\ref{lo_fprot_corr}, using 
$N\!=\!21$ (solid line) and $N\!=\!33$ (dashed line). 
Although the correlations 
decay somewhat faster in the model, the results are very similar to 
those for the real proteins. 

\subsection{Global Structure and Stability}
It is an essential feature of any plausible model for protein folding that it allows for 
structural stability. 
In the following section, we examine our model with respect to this important aspect 
using two methods.
First, we  study structural stability by measuring mean square fluctuations, 
$\langle \delta^2 \rangle$. Second, the structure of the folded conformations will be characterized
by using a measure for $\alpha$-helix formation.

A commmonly used measure of the similarity between two conformations $a$ and $b$
is the mean square distance
 $\delta^2_{ab}$, which is defined as
\begin{equation}
\delta^2_{ab} = \min {1\over N}\sum_{i=1}^N 
|\bar x^{(a)}_i-\bar x^{(b)}_i|^2
\label{d2_equation}
\end{equation}
where $|\bar x^{(a)}_i-\bar x^{(b)}_i|$ denotes the distance 
between the sites $\bar x^{(a)}_i$ and $\bar x^{(b)}_i$, 
and where the minimum is taken over translations and rotations.
The probability distribution $P(\delta^2)$, measured on a thermodynamic ensemble 
with fixed sequence and fixed temperature,
gives valuable information
about structural stability~\cite{Iori}. In particular, if there is a number of 
relatively well-defined structures present,
the distribution will have
a narrow peak close to $\delta^2\approx 0$;
in this case the ensemble contains many pairs of similar conformations.

\begin{figure}[tb]
\mbox{
\epsfxsize=15cm
\epsfbox{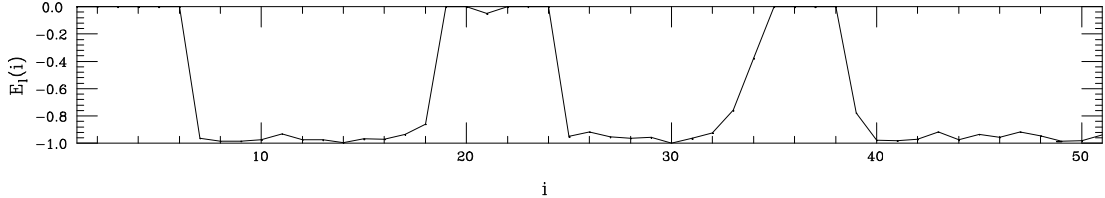}
}
\caption{ 
The function $E_l(i)$ of Eq.~\ref{local_equation}
for the experimental structure of DNA binding protein, PDB accession code 1enh ($N\!=\!54$). 
One easily identifies three $\alpha$-helices from this plot; for $\alpha$-helical structure
$E_l(i)$ is close to minus one.
The resulting $\alpha$-helix assignments are in agreement with those in the PDB entry.
} 
\label{pdb_local}
\end{figure}

Our second method for characterizing structure and stability is to monitor the formation of $\alpha$-helical structure.
A convenient way of doing this is to utilize the function
$E_l(i)$ (Eq.~\ref{local_equation});
besides using it for modelling hydrogen-bonding in the $\alpha$-helical structure,
$E_l(i)$ can also be used to detect $\alpha$-helices in a given structure. 
To illustrate
this, we plot $E_l(i)$ for the experimental structure of the DNA binding protein 1enh
in Fig.~\ref{pdb_local}. From this plot, it is immediately clear
that the structure contains three $\alpha$-helices. The first $\alpha$-helix is identified by the drop
of $E_l(i)$ from zero to slightly above minus one in the region from position $i\!=\!7$ to $i\!=\!19$. 
It should be observed that 16 amino acids participate in this helix; the positions of four
successive $C_{\alpha}$-atoms determine one value of $E_l(i)$.
Below, we will perform a similar analysis for our model proteins.

For the $N\!=\!21$ sequence at $T\!=\!0.15$, the distribution $P(\delta^2)$ has a mean value of
$\langle \delta^2 \rangle=3.6 \pm 0.5\mbox{\AA}^2$. This result may be compared to those of~\cite{Irback:97} 
for the AB model. There, six different $N\!=20$ sequences were studied, and $\langle \delta^2 \rangle$ 
varied between $1\mbox{\AA}^2$ and $10\mbox{\AA}^2$. 
It should be pointed out, however, that the size of the monomers, as measured by the dimensionless ratio
$A_{ij}/r_{i,i+1}$, is slightly larger in the present model (it is equal to one in the AB model).

\begin{center}
\begin{figure}
\mbox{
\epsfxsize=4cm
\epsfbox{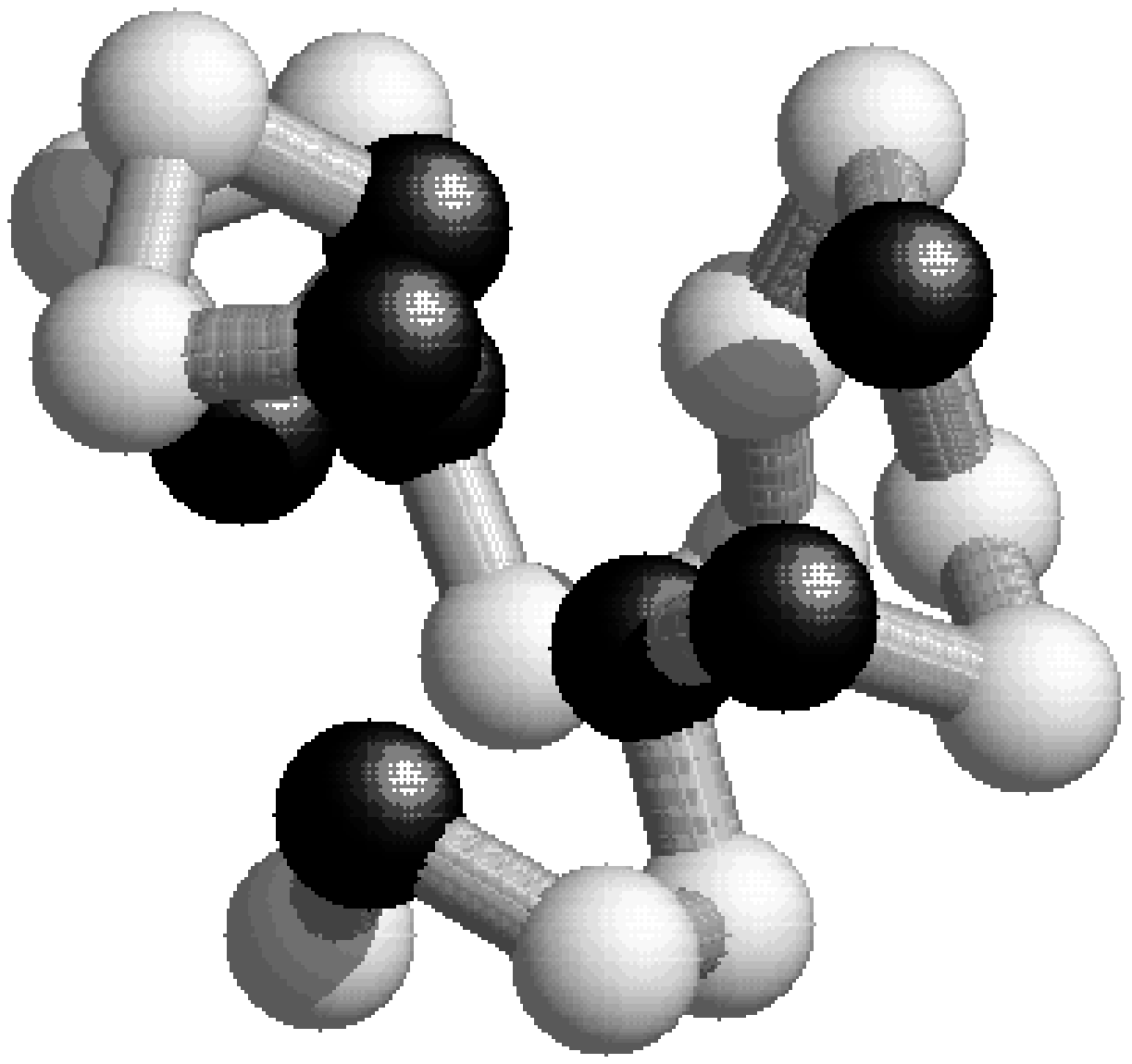}
\epsfxsize=4cm
\epsfbox{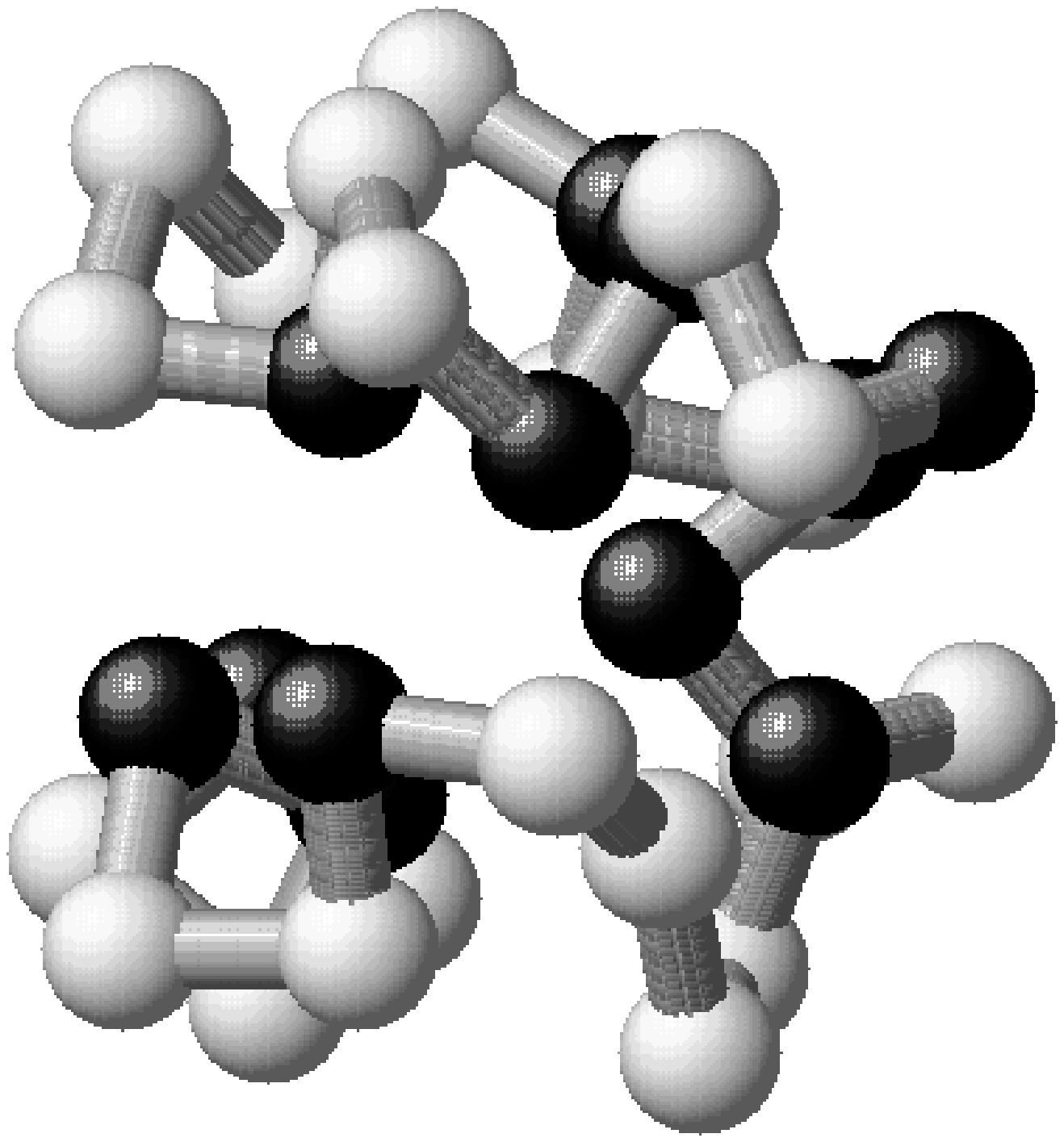}
\epsfxsize=6.5cm
\epsfbox{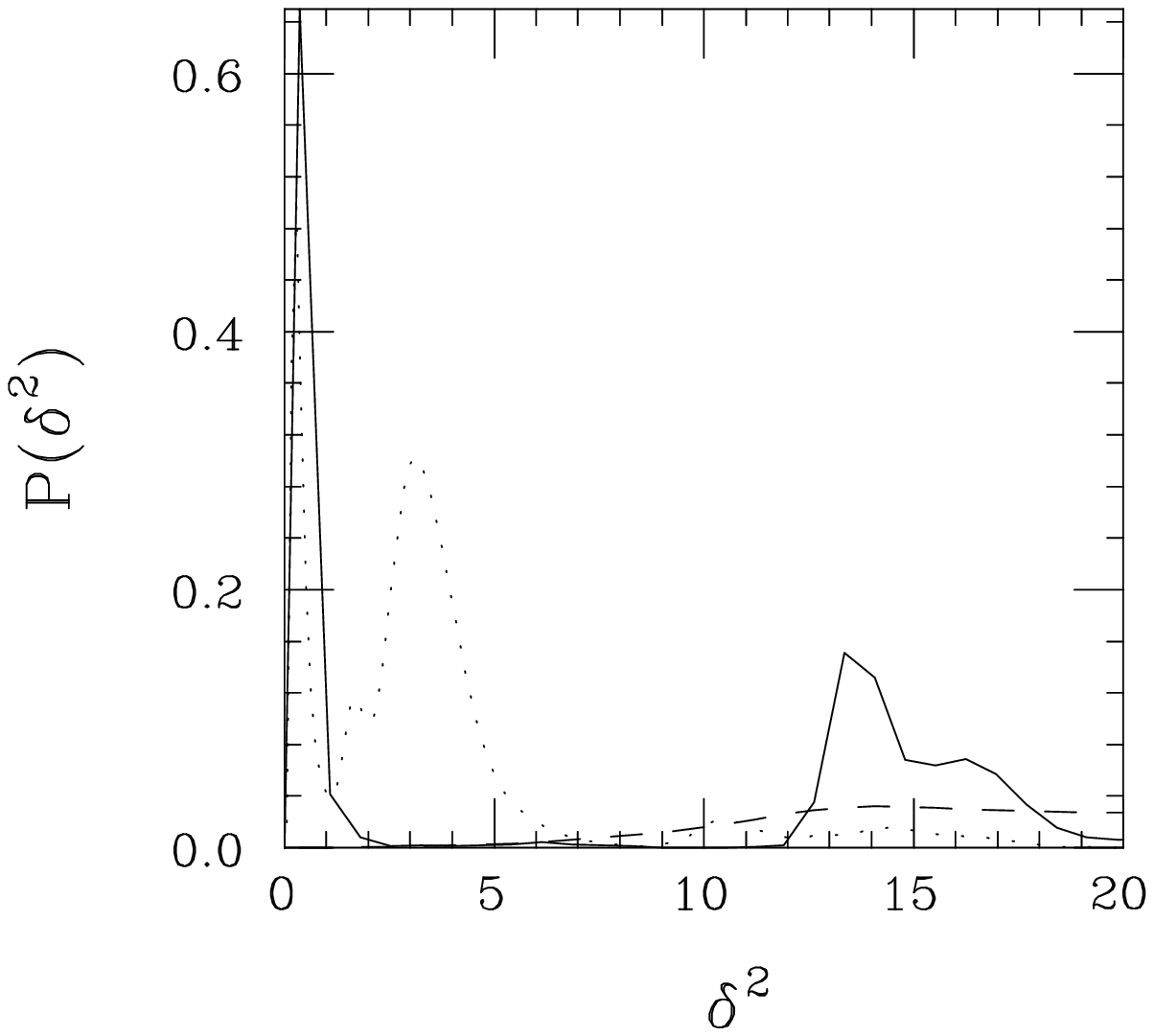}
}
\caption{Representative structures for the most probable states for 
$N\!=\!21$ (left) and $N\!=\!33$ (middle).
Hydrophobic and hydrophilic amino acids are represented by black and white spheres, respectively.
%If the reader manages with her right hand to position the tip of her index finger orthogonal ontop 
%of her thumb (in the difficult orientation),
%she obtains the same overall topology as the displayed  $N\!=\!33$ structure (point your thumb into the most visible
%helix. The black residue on the top right is the tip of your index finger).
On the right side, $P(\delta^2)$ is plotted for (1) $N\!=\!21$, $T\!=\!0.15$ (dotted line), (2) 
$N\!=\!33$, $T\!=\!0.15$ (solid line), and (3) $N\!=\!33$, $T\!\approx\!0.40$ (dashed line).
}
\label{lowstates}
\end{figure}
\end{center}

\begin{figure}[tb]
\mbox{
\epsfxsize=15cm
\epsfbox{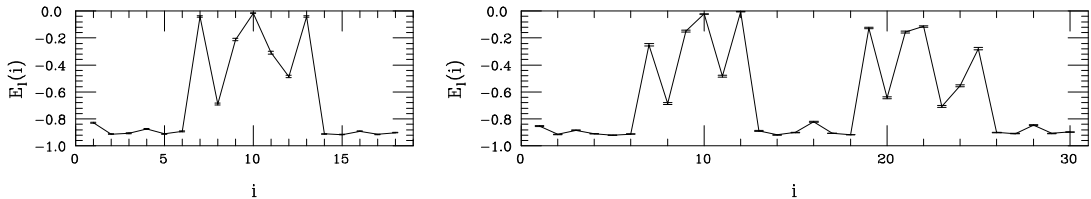}
}
\caption{ Thermodynamic averages of the local interactions $ E_l(i)$ at $T=0.15$
for  $N\!=\!21$ (left) and $N\!=\!33$ (right).
These curves clearly show that the dominating states for $N\!=\!21$ and $N\!=\!33$ contain
2 and 3 helices, respectively. This is also supported by the fact that
fluctuations of bend and torsional angles are small in 
the helical regions.
} 
\label{model_local}
\end{figure}

One could be tempted to conclude from the $P(\delta^2)$ distribution, which is shown in Fig.~\ref{lowstates} (right, dotted line),
that the $N\!=\!21$ system has two similar but distinct states.
A closer analysis reveals that there are actually four distinct states; the system spends almost all time, 95\%,
in the near vincinities of these. This can be shown by measuring the simultaneous $\delta^2$ distances
to all these four states. The mutual distances between these states range from 
1.5$\mbox{\AA}^2$ to 3.7$\mbox{\AA}^2$. 
When it comes to helix content, these four states are very similar. The function $E_l(i)$, 
as displayed in Fig.~\ref{model_local},
makes it evident that the overall structure is that of a two helix bundle. We also looked at the expectation values
of the bend and torsional angles and their variances, which showed that the helices are indeed very stable and 
common to the four different states.  The structure of one of the four states is shown in Fig.~\ref{lowstates} (the
other three are very similar). From Fig.~\ref{lowstates} it can also be seen that (1) we succeed to obtain a hydrophobic
core and (2) the $\alpha$-helices show the patterning intended in the construction of the sequences. 

The $P(\delta^2)$ distribution for $N\!=\!33$, see Fig.~\ref{lowstates} (right plot), 
also has a pronounced peak at $\delta^2\approx 0$.
However, a significant amount
of probability is also found around $\delta^2\approx 15 \mbox{\AA}^2$.
The existence of this outlier in  $P(\delta^2)$ shows that the overall
geometry must be different for the different states present. 
A detailed analysis, similar to that for  $N\!=\!21$,  shows that 
$P(\delta^2)$ is dominated by three states having
mutual distances of around $13.8, 15.0$  and $16.3\mbox{\AA}^2$. Taken together, these three 
states contain 97\% probability at $T\!=\!0.15$, distributed as 68\%, 17\% and 12\%. 
These three states share three common $\alpha$-helices; the positions of the helices along
the chain can be seen from Fig.~\ref{model_local}. 
How do the three states differ? For the most probable state (68\%),  
the three helices are coordinated in such a way that the overall topology has 
right-handed chirality , 
as can be seen from Fig.~\ref{lowstates}, middle plot. For this state, 26 monomers participate 
in the three helices.
The other two states are described as follows: 
the first two helices and the connecting loop
form a U-shaped entity (the helices are not really parallel). 
The third helix points back into the U, for one state on top and for the other
below the U-plane. The right-handed state is slightly less probable (12\%) 
than the left-handed one (17\%). 
The helices of the most probable state are present in
these two states as well; however, the helices of these two states are longer with 31 and 33 monomers
participating in the sense that 22 and 24 values of $E_l(i)$ are close to minus one.

Knowing about this three-fold structural degeneracy could be of interest for protein design.
If the design is done in terms of the hydrophobicity pattern, as for example in~\cite{Kamtekar},
the designer could easily knock out two unwanted states by introducing unfavorable
residues.

As stated in section 2.3, we also tried to simulate a sequence with  $N\!=\!45$, 
intended to design 4-helix topology. 1000 CPU hours were insufficient to
obtain fully reliable results. However, we feel confident that we managed
to adjust the weights $g_k$ pretty well and obtained a selection of low-energy
structures that we rate as representative when it comes to local structure and
overall size of the chain. To give an idea about the size of our model chains,
we display the $C_{\alpha}$-$C_{\alpha}$ distance distribution function in Fig.~\ref{lo_fprot_corr}
(right figure, dashed line). In this graph, we compare it with the same distribution 
for the DNA binding protein 1enh, which contains $54$ amino acids. The general
shape of the two curves is similar, taking into account the shorter length of the model
protein. This shows that even the size of our model proteins resembles that of real ones.

\section{Conclusions}
A minimal off-lattice model for $\alpha$-helical proteins has been presented. 
$\alpha$-helical structure is obtained by  setting local LJ parameters
to values consistent with  $\alpha$-helical geometry and introducing local interactions
modelling chirality and $\alpha$-helical hydrogen bonding.  
The global LJ parameters are determined by the size of an average amino acid.

Thermodynamic results for two sequences of length $21$ and $33$ have been presented. 
The construction of these sequences was inspired by
the method of Kamtekar~\cite{Kamtekar} to design real de novo proteins.
This turns out to be a successful strategy here as well, since
we observe $\alpha$-helices at the intended positions along the chains.
Furthermore, the helices observed in our model show the intended hydrophobicity
pattern with one side being hydrophobic and the other hydrophilic;
the hydrophobic sides of the helices point inwards.

The local structure has been explored using angle distributions and bond-bond correlations.
When we compare results from our model with those for real $\alpha$-helical proteins,
these observables are very similar. 

By inspecting the mean square distribution, $P(\delta^2)$, it has
convincingly be shown that the $N\!=\!21$ sequence is structurally stable. The overall
topology was found to be that of a 2-helix bundle. For $N\!=\!33$, we found three
globally different states which dominate the low temperature phase. 
These three states have three $\alpha$-helices
in common and differ from each other by the orientation of these helices. 

Simulated tempering is the key to our ability investigate the low-temperature thermodynamics.
The simulations took 30 CPU hours for $N\!=\!21$  and 500 CPU hours for  $N\!=\!33$. 
1000 CPU hours were not sufficient to obtain reliable results for $N\!=\!45$, but
we feel confident that we managed to properly adjust the weights $g_k$. 
Therefore, we believe that simulating
the $N\!=\!45$ sequence in a reliable way will be possible in the near future.
Also, we feel confident that the simulations can be speeded up by further algorithmic improvements,
such as a better way of distributing the temperatures $T_k$ (see e.g.~\cite{Hansmann}).

\section{Appendix}
\subsection{The Ideal $\alpha$-Helix: Model Parameters}
Five parameters of our model are determined by the geometry of an ideal $\alpha$-helix.
A sixth parameter is determined by the average size of an amino acid.
In this section, the calculation of these parameters is presented.

\begin{figure}
\begin{center}
\mbox{
\epsfxsize=4.5cm
\epsfbox{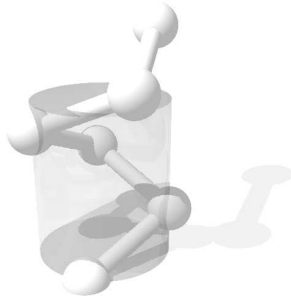}
}
\caption{Illustration of an ideal $\alpha$-helix: The polypeptide backbone
has 3.6 amino acids per turn  with a pitch of $5.4\mbox{\AA}$, as indicated
by the height of the cylinder. The cylinder has a diameter of $4.5..\mbox{\AA}$. 
}
\label{helixpic}
\end{center}
\end{figure}

A discrete helix is uniquely described by its bend and torsional angles,
or, equivalently, by its pitch, the number of entities per turn and its chirality.
An ideal $\alpha$-helix has a translation of 5.4$\mbox{\AA}$ and 3.6 amino acids
per turn with right-handed chirality, as is illustrated in Fig.~\ref{helixpic}.
This is easily reformulated in terms of bend and torsional angles
which we denote by $\hat{\tau}$ and $\hat{\alpha}$, respectively. 

In order to make everything well-defined, we have to give an exact definition of
bend and torsional angles.
Three succesive bond vectors $\vec{b}_i,\vec{b}_{i+1},\vec{b}_{i+2}$ define the
two bend angles $\tau_i$,$\tau_{i+1}$ and the torsional angle $\alpha_{i}$.
The bend angle $\tau$ is defined to be $180^{\circ}$ if the two involved bond vectors are parallel.
The torsional angle $\alpha$ is taken to be the angle between the two vectors
$\vec{b}_i \times \vec{b}_{i+1}$ and $\vec{b}_{i+1} \times \vec{b}_{i+2}$.
Therefore, $\alpha\!=\!0$ if and only if the three involved bond vectors are
situated in a plane.
 $\alpha$ ranges from
$-180^{\circ}$ to $180^{\circ}$, and is chosen to be negative for a right-handed and positive 
for a left-handed system. 

A single LJ term in Eq.~\ref{energy_definition},
\begin{equation}
4\epsilon\left(  \left(\frac{{A_{ij}}}{{r_{ij}}}\right)   ^{12}-
\left(   \frac{A_{ij}}{r_{ij}}   \right)^{6}\right)\,,
\end{equation}
has its minimum of depth $\epsilon$ at $r_{ij}=A_{ij}\cdot 2^{1/6}$; the minimum can be moved
by changing $A_{ij}$. We adjust
$A_{ij}$ for $\mid\! i-j\!\mid \leq 4$ such that the minimum is situated at the distance observed in 
an ideal $\alpha$-helix, $r_{ij}^{\alpha}$. 
For all other pairs,  
$\mid\!j\!-\!i\mid>4$, we choose
the LJ parameter $A_{ij}$ by tuning the volume of the  LJ sphere to that of an average
amino acid, $161\mbox{{\AA}}^3$:
\beq
\mid\!i\!-\!j\!\mid >4 \;\Rightarrow\; A_{ij} = 2^{-1/6} \times  \sqrt[3]{\frac{3}{4\pi}161{\mbox{\AA}^3}}
\approx 6.01{\mbox{\AA}}
\eeq

To summarize, one obtains the following values for $\hat{\tau}$, $\hat{\alpha}$, $r_{ij}^\alpha$ and $A_{ij}$:
\begin{center}
$\hat{\tau}\!\approx\!90.53^{\circ}\!\;\;\;\;\;\;$   
$\hat{\alpha}\!\approx\!-129.61^{\circ}\!$ \\ %-2.2621681534.. 
\vspace{0.1cm}
\begin{tabular}{rccccc}
                              & $\mid\!j\!-\!i\!\mid=1$ & $\mid\!j\!-\!i\!\mid=2$ &
                                $\mid\!j\!-\!i\!\mid=3$ & $\mid\!j\!-\!i\!\mid=4$  & $\mid\!j\!-\!i\mid>4$ \\ \hline
$r_{ij}^\alpha$ [$\mbox{\AA}$]&     3.8     & 5.39.. & 5.04.. & 6.19.. &    -               \\
%$A_{ij}$   [$\mbox{L}$]       &     -       & 1.26.. & 1.18.. & 1.45.. &  1.58..    \\  
$A_{ij}$   [$\mbox{\AA}$]     &     -       & 4.80.. & 4.49.. & 5.52.. &  6.01..    \\  
\end{tabular}
\end{center} 

\subsection{The AB model}
The $\alpha$-helical model presented in this paper is derived from the 3D AB
model suggested recently by Irb\"ack et al.~\cite{Irback:97}, which in turn is closely related to
the 2D model proposed by Stillinger et. al~\cite{Stillinger,Irback:95}.
To make comparisons easier, we here give the energy functions for both the AB model and the present model;

\begin{eqnarray}
E(\vec{b}; \sigma) &=&    
\sum_{i=1}^{N-2}\sum_{j=i+2}^N 
4\epsilon(\sigma_i,\sigma_j)\left(  \left(\frac{A_{ij}}{{r_{ij}}}\right)   ^{12}-
\left(   \frac{A_{ij}}{r_{ij}}   \right)^{6}\right)\ 
-\kappa_1\sum_{i=1}^{N-2}\frac{\vec{b}_i\cdot\vec{b}_{i+1}}{|\vec{b}_i||\vec{b}_{i+1}|}  
-\kappa_2\sum_{i=1}^{N-3}\frac{\vec{b}_i\cdot\vec{b}_{i+2}}{|\vec{b}_i||\vec{b}_{i+2}|}\nonumber \\ 
E(\vec{b}; \sigma) &=&  
\sum_{i=1}^{N-2}\sum_{j=i+2}^N 
4\epsilon(\sigma_i,\sigma_j)\left(  \left(\frac{{A_{ij}}}{{r_{ij}}}\right)   ^{12}-
\left(   \frac{A_{ij}}{r_{ij}}   \right)^{6}\right)\ - \sum_{i=1}^{N-3} \exp\left( -\! \frac{(\alpha_i\!-\!\hat{\alpha})^2\!+\!(\tau_i\!-\!\hat{\tau})^2\!
+\!(\tau_{i+1}\!-\!\hat{\tau})^2}{2 w^2}\right) \nonumber
\end{eqnarray}

They differ only in the choice of $A_{ij}$ and in the local interactions; for the AB model, $A_{ij}=3.8\mbox{{\AA}}$ independent of $i$ and $j$.
The dependence of the AB model on the strengths of the local interactions $(\kappa_1,\kappa_2)$
was discussed to some extent in~\cite{Irback:97}, where it was found that the local interactions
are necessary in order to obtain regularities in the local structure and thermodynamic stable sequences.
The final choice of  $(\kappa_1,\kappa_2)$ in~\cite{Irback:97} was   $(\kappa_1,\kappa_2)=(-1,0.5)$.

\section{Acknowledgements}
This project is supported by the ``Swedish Foundation for Strategic Research''.
Thanks to Anders Irb{\"a}ck and Carsten Peterson for useful discussions.
I appreciate helpful comments on the manuscript by Anders Irb\"ack, Robert van Leeuwen and Peter Sutton.
This project would not have been started without the
encouragement from Conrad Newton.


\begin{thebibliography}{99}

\bibitem{Karplus:95}
See e.g. M. Karplus and A. \u{S}ali,
``Theoretical Studies of Protein Folding and Unfolding'',
\COSB {\bf 5}, 58 (1995).

\bibitem{Lau:89}
K.F. Lau and K.A. Dill,
``A Lattice Statistical Mechanics Model of the Conformational and
Sequence Spaces of Proteins'',
\Mac {\bf 22}, 3986 (1989).

%\bibitem{Chan:91}
%H.S. Chan and K.A. Dill,
%```Sequence Space Soup' of Proteins and Copolymers'',
%\JCP {\bf 95}, 3775 (1991).

\bibitem{Camacho:93}
C.J. Camacho and D. Thirumalai,
``Minimum Energy Compact Structures of Random Sequences of Heteropolymers'',
\PRL {\bf 71}, 2505 (1993).

%\bibitem{Chan:94}
%H.S. Chan and K.A. Dill,
%``Transistion States and Folding Dynamics of Proteins and Heteropolymers'',
%\JCP {\bf 100}, 9238 (1994).

\bibitem{Dill:95}
K.A. Dill, S. Bromberg, K. Yue, K.M. Fiebig, D.P. Yee, P.D. Thomas and 
H.S. Chan,
``Principles of Protein Folding --- A Perspective from Simple Exact Models'',
\ProSci {\bf 4}, 561 (1995).

\bibitem{Erik}
A. Irb\"ack and E. Sandelin,
``Local Interactions and Protein Folding: A Model Study on the Square and Triangular Lattices'',
\JCP {\bf 108}, 2245 (1998).

\bibitem{Iori}G. Iori, E. Marinari and G. Parisi, 
``Random Self-Interacting Chains: A Mechanism for Protein Folding'',
\JP {\bf A} {\bf 24}, 5349 (1991).

\bibitem{Fukugita:93}
M. Fukugita, D. Lancaster and M.G. Mitchard,
``Kinematics and Thermodynammics of a Folding Heteropolymer'',
\PNAS {\bf 90}, 6365 (1993).

\bibitem{Veitshans}
T. Veitshans, D.K. Klimov and D. Thirumalai,
``Protein Folding Kinetics: Time Scales, Pathways, and Energy Landscapes
in Terms of Sequence Dependent Properties'',
\FD {\bf 2}, 1 (1997).

\bibitem{Stillinger} F.H. Stillinger, T. Head-Gordon and C.L. Hirschfeld,
``Toy Model for Protein Folding'', 
\PR {\bf E48}, 1469 (1993).

\bibitem{Irback:95}
A. Irb\"ack and F. Potthast,
``Studies of an Off-Lattice Model for Protein Folding: Sequence
  Dependence and Improved Sampling at Finite Temperature'',
\JCP {\bf 103}, 10298 (1995).

\bibitem{Irback:97}
A. Irb\"ack, C. Peterson, F. Potthast and O. Sommelius,
``Local Interactions and Protein Folding: A 3D Off-Lattice Approach'',
\JCP {\bf 107}, 273 (1997).

%\bibitem{Pauling}
%L. Pauling,
%``The Structure of Proteins: Two hydrogen-bonded Helical Conformations of the Polypeptide Chain'',
%\PNAS {\bf 37}, 205 (1951).

\bibitem{Rey}
A. Rey and J. Skolnick,
``Computer Modeling and Folding of Four-Helix Bundles'',
\Pro {\bf 16}, 8 (1993).


\bibitem{Kamtekar}
S. Kamtekar, J.M. Schiffer, H. Xiong, J.M. Babik and M.H. Hecht,
``Protein Design by binary Patterning of polar and nonpolar Amino Acids'',
{\it Science} {\bf 262}, 1680 (1993). 

\bibitem{Creighton:proteins}
T. Creighton, 
{\it Proteins: Structures and Molecular Properties} (Freeman, New York, 1993). 

\bibitem{West}
M.W. West and M.H. Hecht,
``Binary Patterning of Polar and Nonpolar Amino Acids in the Sequences and
Structures of Native Proteins'',
\ProSci {\bf 4}, 2032 (1995).

\bibitem{Ryssland}
A.P. Lyubartsev, A.A. Martsinovski, S.V. Shevkunov and 
P.N. Vorontsov-Velyaminov,
``New Method to Monte Carlo Calculation of the Free Energy:
Method of Expanded Ensembles'',
\JCP {\bf 93}, 1776 (1992).

\bibitem{Marinari:92}E. Marinari and G. Parisi, 
``Simulated Tempering: A New Monte Carlo Scheme'',
{\em Europhys. Lett. \bf 19}, 451 (1992).

%http://scop.mrc-lmb.cam.ac.uk/scop/
\bibitem{scop}
A.G. Murzin, S.E. Brenner, T. Hubbard and C. Chothia,
"SCOP: A Structural Classification of Proteins Database for the Investigation of Sequences and Structures",
\JMB {\bf 247}, 536 (1995).

\bibitem{lo_Bernstein:77}
F.C. Bernstein, T.F. Koetzle, G.J.B. Williams, E.F. Meyer, M.D. Brice,
  J.R. Rodgers, O. Kennard, T. Shimanouchi  and M. Tasumi,
``The Protein Data Bank: A Computer Based Archival File for
  Macromolecular Structures'',
\JMB {\bf 112}, 535 (1977).

\bibitem{KISSINGER}
C.R. Kissinger, B.S. Liu, E. Martin-Blanco, T.B. Kornberg and C.O. Pabo,
``Crystal Structure of an engrailed Homeodomain-DNA Complex at 2.8 Resolution: A Framework for
understanding Homeodomain-DNA Interactions'',
{\it Cell} {\bf 63}, 579 (1990).  

\bibitem{Hansmann}
U.H.E. Hansmann,
``Parallel Tempering Algorithm for Conformational Studies of Biological Molecules'',
{\it Chem.\ Phys.\ Lett.\ } {\bf 281}, 140 (1997).
%%%%%%%%%%%%%%%%%%%%%%%%%%%%%%%%%%%%%%%%%%%%%%%%%%%%%%%%%%%%%%%%%%%%%%%%%%


%\bibitem{Chan} H.S. Chan and K.A. Dill,
%``Origins of Structure in globular Proteins'',
%\PNAS {\bf 87} 6388-6392 (1990).

%\bibitem{Reese} M.G. Reese, O.Lund, J.Bohr, H.Bohr, J.E. Hansen, and S. Brunak, 
%``Distance distributions in proteins: A six parameter representation'', 
%{\it Protein Engineering}, {\bf 9}, 733-740  (1996). 

%\bibitem{Abkevich:95}V.I. Abkevich, A.M. Gutin and E.I. Shakhnovich,
%``Impact of Local and Non-Local Interactions on Thermodynamics and Kinetics of 
%Protein Folding'',
%{\it Journal of Molecular Biology} {\bf 252} 460-471 (1995).

%\bibitem{lo_Socci:94a}
%N.D. Socci, W.S. Bialek and J.N. Onuchic,
%``Properties and Origins of Protein Secondary Structure'',
%\PR {\bf E 49}, 3440 (1994).

%\bibitem{Gregoret:91}
%L.M. Gregoret and F.E. Cohen,
%``Protein Folding. Effect of Packing Density on Chain Conformation'',
%\JMB {\bf 219}, 109-122 (1991).


%\bibitem{Berriz}
%G.F. Berriz, A.M.Gutin, and E.I.Shakhnovich, 
%``Cooperativity and Stability in a Langevin Model of Protein Folding'',
%\JCP (1997).

%\bibitem{Aszodi}
%A. Aszodi and W.R. Taylor,
%`` Secondary structure formation in model polypeptide chains''
%{\it Protein Engineering} {\bf 7}, 633-644 (1994).
%still not available.....get it !

%\bibitem{Kirkpatrick}
%S. Kirkpatrick, C.D. Gelatt and M.P. Vecchi,
%``Optimization by Simulated Annealing'',
%{\it Science} {\bf 220} 671 (1983).

%\bibitem{Cohen}
%F.E. Cohen and M.J.E. Sternberg,
%``On the prediction of protein structure: The significance of the 
%root-mean-square deviation'',
%\JMB {\bf 138} 321-333 (1980).

%\bibitem{Bastolla}
%U. Bastolla and P.Grassberger,
%``Phase Transitions of Single Semi-stiff Polymer Chains''
%{\it cond-mat/9705178} (1997).

%\bibitem{Doniach}
%S. Doniach, T. Garel and H. Orland,
%``'Phase Diagram of a semiflexible polymer chain in a $\theta$
%solvent: Application to protein folding'',
%\JCP {\bf 105} 1601-1608 (1996).

%\bibitem{Karplus}
%Y. Zhou, M. Karplus, J.M. Wichert and C.K. Hall,
%``Equilibrium thermodynamics of homopolymers and clusters:
%Molecular dynamics and Monte Carlo simulations of systems with 
%square-well interactions'',
%\JCP {\bf 107} 10691-10708 (1997).

%\bibitem{Baumgartner}
%A. Baumg\"artner,
%``Statics and dynamics of the freely jointed polymer chain with Lennard-Jones interaction''
%\JCP {\bf 72} 871-879



%\bibitem{lo_Hobohm}
%U. Hobohm and C. Sander,
%``Enlarged Representative Set of Protein Structures'',
%\ProSci {\bf 3}, 522 (1994).
\end{thebibliography}
\end{document}